\documentstyle[aps,prd]{revtex}
\catcode`\@=11 \@addtoreset{equation}{section}
\catcode`\@=11

\draft
\begin{document}

\twocolumn[\hsize\textwidth\columnwidth\hsize
\csname @twocolumnfalse\endcsname

\title{Thermal properties of spacetime foam}
\author{Luis J. Garay}
\address{Instituto de Matem\'{a}ticas y F\'{\i}sica Fundamental,
CSIC, C/ Serrano 121, 28006 Madrid, Spain}
\date{August 25, 1998}
\maketitle

\begin{abstract}

Spacetime foam can be modeled in terms of nonlocal effective
interactions in a classical nonfluctuating background. Then, the
density matrix for the low-energy fields evolves, in the
weak-coupling approximation, according to a master equation that
contains a diffusion term. Furthermore, it is argued that
spacetime foam behaves as a quantum thermal field that, apart
from inducing loss of coherence, gives rise to effects such as
gravitational Lamb and Stark shifts as well as quantum damping in
the evolution of the low-energy observables. These effects can
be, at least in principle, experimentally tested.

\end{abstract}

\pacs{PACS: 04.60.-m, 03.65.Bz, 04.20.Gz, 04.70.Dy
\hfill {\it gr-qc/9806047; Phys. Rev. D58, 124015 (1998)}}
 ]
\footnotetext{Copyright 1998 by The American Physical Society}

\section{Introduction}

It seems natural to assume that spacetime at the Planck scale
must have a very complicated and ever-changing topology. Indeed,
it was Wheeler \cite{wheeler} who first suggested the foamlike
structure of spacetime \cite{wheeler,ha78,ca97} as an inescapable
ingredient of the yet-to-be-built quantum theory of gravity.
Since then, various spacetime foam components have been proposed:
wormholes \cite{haw88,col88}, virtual black holes \cite{96haw01},
and quantum time machines \cite{ha95,pedro} among them.

The quantum theory of gravity suffers from problems \cite{is93}
that have remained unsolved for many years. They originate in the
fact that gravity deals with the frame in which everything takes
place, i.e., with spacetime, in sharp contrast with any other
interaction, for which spacetime is a passive frame. When gravity
is brought onto the scene, the frame itself becomes dynamical. It
suffers the quantum fluctuations of the other interactions and,
even more, introduces its own fluctuations, thus becoming an
active agent in the theory.

We are used to putting everything into spacetime, so that we can
name and handle events. General relativity made spacetime alive
and in this sense, was a major change. But, although dynamical,
the relations between different events were still sharply
defined. Quantum mechanics changed this, too. In such a dynamical
frame, objects became fuzzy; exact locations were substituted by
probability amplitudes of finding an object in a given region of
space at a given instant of time.

A quantum uncertainty in the position of a particle implies an
uncertainty in its momentum and, therefore, due to the
gravity-energy universal interaction, would also imply an
uncertainty in the geometry, which in turn would introduce an
additional uncertainty in position of the particle. The geometry
would thus be subject to quantum fluctuations that would
constitute the spacetime foam and that should be of the same
order as the geometry itself at the Planck scale. This would give
rise to a minimum length \cite{95gar01} beyond which the
geometrical properties of spacetime would be lost, while on
larger scales it would look smooth and with a well-defined metric
structure.

The quantum structure of spacetime would be relevant at energies
close to Planck scale and one could expect that the quantum
gravitational virtual processes that constitute the spacetime
foam could not be described without knowing the details of the
theory of quantum gravity. However, the gravitational nature of
spacetime fluctuations provides a mechanism for studying the
effects of these virtual processes in low-energy physics. Indeed,
virtual gravitational collapse and topology change would forbid a
proper definition of time at the Planck scale. More explicitly,
in the presence of horizons, closed timelike curves, topology
changes, etc., any Hamiltonian vector field that represents time
evolution outside the fluctuation would vanish at points inside
the fluctuation. This means that it would not be possible to
describe the evolution by means of a Hamiltonian unitary flow
from an initial to a final state and, consequently, quantum
coherence would be lost. These effects and their order of
magnitude would not depend on the detailed structure of the
fluctuations but rather on their existence and global properties.
In general, the regions in which the asymptotically timelike
Hamiltonian vector fields vanish are associated with infinite
redshift surfaces and, consequently, these small spacetime
regions would behave as magnifiers of Planck length scales
transforming them into low-energy modes as seen from outside the
fluctuations \cite{pa98}. Therefore, spacetime foam and the
related minimum length would affect low-energy physics, so that
low-energy experiments would effectively suffer a nonvanishing
uncertainty. In this situation, a loss of quantum coherence would
be almost unavoidable \cite{unpred}. In fact, Hawking
\cite{96haw01} has pointed out that scalar fields may lose
coherence extremely fast and that the loss of quantum coherence
might also be responsible for the vanishing of the $\theta$ angle
of quantum chromodynamics.

In this paper, we show that spacetime foam behaves as a quantum
thermal bath with a nearly Planckian temperature that has a weak
interaction with low-energy fields. As a consequence, other
effects, apart from a loss of coherence, such as Lamb and Stark
transition-frequency shifts, quantum damping, and cold diffusion,
characteristic of systems in a quantum environment
\cite{91gar01,car93}, naturally appear as low-energy predictions
of this model. A brief account of these results has already
appeared in Ref.~\cite{qfoam}. This kind of quantum gravitational
effects can be, in principle, experimentally tested (see, e.g.,
Refs. \cite{el84,am98}), as we also argue in this work.

This paper is organized as follows. In Sec. \ref{effint}, we
propose an effective model of spacetime foam in terms of nonlocal
interactions and argue that it is equivalent to a local theory
with a stochastic classical Gaussian noise source. In Sec.
\ref{class}, a master equation for the low-energy density matrix
is obtained and the diffusion term is studied. Section \ref{qme}
is devoted to the quantum effects that have not been taken into
account in previous sections and derive a master equation that
includes them. It is also shown that spacetime foam can be
described as a quantum thermal bath and the consequences of this
effective behavior are analyzed. We close this section with a
short discussion about the kind of experiments and observations
that could be sensitive enough to test these effects. In Sec.
\ref{compo}, we study the role that some of the components of
spacetime foam (wormholes, virtual black holes and quantum time
machines) play in the effective theory. We summarize and conclude
in Sec. \ref{concl}.

\section{Effective interactions}
\label{effint}

In this section, we will construct an effective theory for the
evolution of low-energy fields in spacetime foam, where we
possibly have a finite resolution limit because the notion of
distance is not valid at the quantum gravitational scale.

With this aim, we will substitute spacetime foam by a fixed
classical nonfluctuating background with low-energy fields living
on it. We will perform a 3+1 foliation of the effective spacetime
that, for simplicity, will be regarded as flat, $t$ denoting the
time parameter and $x$ the spatial coordinates. Spacetime foam
features, i.e., the gravitational fluctuations and the minimum
length generated by them, will be characterized by nonlocal
interactions. They will relate spacetime points that are
sufficiently close in the effective nonfluctuating background,
where a well-defined notion of distance exists. These effective
nonlocal interactions will be described in terms of local
interactions as follows.

Let us consider a basis $\{h_i(t)\}$ of local gauge-invariant
interactions at the spacetime point $(x,t)$, each element
consisting of factors of the form
$l_*^{2n(1+s)-4}\left[\phi(x,t)\right]^{2n}$, and $\phi$ being
the low-energy field strength of spin $s$. As a notational
convention, each index $i$ implies a dependence on the spatial
position $x$ by default; whenever the index $i$ does not carry an
implicit spatial dependence, it will appear underlined
${\underline{i}}$. Also, any contraction of indices (except for
underlined ones) will entail an integral over spatial positions.
Then, the nonlocal effective interaction can be included in the
Euclidean action by means of a term of the form
\begin{equation}
I_{\rm int}=\sum_N {\cal I}_N ,
\end{equation}
where $I_{\rm int}$ is the $N$-local interaction term,
\begin{equation}
{\cal I}_N=\frac{1}{N!}\int \! dt_1\cdots dt_N c^{i_1\cdots
i_N}(t_1\ldots t_N)h_{i_1}(t_1)\cdots h_{i_N}(t_N).
\end{equation}
The dimensionless functions $c^{i_1\cdots i_N}(t_1\ldots t_N)$
cannot depend on the location of the gravitational fluctuation
itself because of conservation of energy and momentum: the
fluctuations do not carry energy, momentum, or gauge charges.
Thus, diffeomorphism invariance is preserved, at least at
low-energy scales, provided that the coefficients $c^{i_1\cdots
i_N}(t_1\ldots t_N)$ only depend on relative positions. This
invariance cannot be expected to hold at the Planck scale as
well. However, this violation of energy-momentum conservation is
safely kept within Planck scale limits \cite{95unr01}, where the
processes will no longer be Markovian.

The coefficients $c^{i_1\cdots i_N}(t_1\ldots t_N)$ must vanish
for relative spacetime distances larger than the length scale $r$
of the gravitational fluctuations. Indeed, if the gravitational
fluctuations are smooth in the sense that they only involve
trivial topologies or contain no horizons, the coefficients
$c^{i_1\cdots i_N}(t_1\ldots t_N)$ will be $N$-point propagators
which, as such, will have infinitely long tails and the size of
the gravitational fluctuations will be effectively infinite. In
other words, we would be dealing with a local theory written in a
nonstandard way. The gravitational origin of these fluctuations
eliminate these long tails because of the presence of
gravitational collapse and topology change. This means that, for
instance, virtual black holes \cite{96haw01} will appear and
disappear and horizons will be present throughout. As Padmanabhan
\cite{pa98} has also argued, horizons induce nonlocal
interactions of finite range since the Planckian degrees of
freedom will be magnified by the horizon (because of an infinite
redshift factor) thus giving rise to low-energy interactions as
seen from outside the gravitational fluctuation. Virtual black
holes represent a kind of components of spacetime foam that
because of the horizons and their nontrivial topology will induce
nonlocal interactions but, most probably, other fluctuations with
complicated topology will warp spacetime in a similar way and the
same magnification process will also take place.

Finally, the coefficients $c^{i_1\cdots i_N}(t_1\ldots t_N)$
will contain a factor $[e^{-S(r)/2}]^N$, $S(r)$ being the
Euclidean action of the gravitational fluctuation, which is of
the order $(r/l_*)^2$.  This is just an expression of the idea
that inside large fluctuations, interactions that involve a
large number of spacetime points are strongly suppressed. As the
size of the fluctuation decreases, the probability for events in
which three or more spacetime points are correlated increases,
in close analogy with the kinetic theory of gases: the higher
the density of molecules in the gas, the more probable is that a
large number of molecules collide at the same point. The
expansion parameter in this example is typically the density of
molecules. In our case, the natural expansion parameter is the
transition amplitude. It is given by the square root of the
two-point transition probability which in the semiclassical
approximation is of the form $e^{-S(r)}$.

A simple calculation shows that
\begin{equation}
{\cal I}_{N}\sim\epsilon^N (r/l)^{2N-4}
\prod_{{\underline{i}}=1}^N (l_*/l)^{2n_{\underline{i}}
(1+s_{\underline{i}})-2},
\end{equation}
where $\epsilon=e^{-S(r)/2}(r/l_*)^2$. Indeed, ${\cal I}_{N}$
contains a factor $[e^{-S(r)/2}]^N$ coming from the coefficient
$c^{i_1\cdots i_N}(t_1\ldots t_N)$, as discussed above; each
interaction $h_i$ provides a factor
$(l_*/l)^{2n_{\underline{i}}(1+s_{\underline{i}})}l_*^{-4}$;
there are also $N$ integrals over spacetime positions, $N-1$ of
which are integrals over relative positions and therefore give a
factor $r^4$ each; and, finally, the integral over the global
spacetime position provides an additional factor $l^4$. The
interaction term ${\cal I}_{N}$ has contributions from three
different length scales, Planck length $l_*$, the size of the
gravitational fluctuations $r$, and the low-energy length scale
$l$, through the ratios $r/l_*$, $r/l$ and $l/l_*$: the factor
$\epsilon^N (r/l)^{2N-4}$ depends only on the first two and is
common to all powers $n$ and spins $s$ while the factors
$(l_*/l)^{2n_{\underline{i}}(1+s_{\underline{i}})-2}$ depend only
on the low-energy scale (in Planck units) and contain the
information about the different kind of interactions involved.

The contributions of the trilocal and higher effective
interactions are, at most, of order $\epsilon^3$. Therefore, in
the weak-coupling approximation, i.e., up to second order in the
expansion parameter $\epsilon$, they can be ignored. On the other
hand, the local terms ${\cal I}_{0}$ and ${\cal I}_{1}$ can be
absorbed in the bare action. Indeed,  the coefficient $c$
appearing in ${\cal I}_{0}$ is constant; the coefficients
$c^i(t)$ in ${\cal I}_{1}$ cannot depend on spacetime positions
because of diffeomorphism invariance and are therefore constant
as well. Consequently, we can write the nonlocal interaction
term in the Euclidean action as the bilocal contribution
\begin{equation}
I_{\rm int} =\frac{1}{2}\int \!dt dt^\prime c^{ij}(t-t^\prime)
h_i(t)h_j(t^\prime),
\end{equation}
where we have renamed $c^{ij}(t,t')$ as $c^{ij}(t-t^\prime)$.
This coefficient is symmetric in the pair of indices $ij$ and
depends on the spatial positions $x_{\underline{i}}$ and
$x_{\underline{j}}$ only through the relative distance
$|x_{\underline{i}}-x_{\underline{j}}|$. It is of order
$e^{-S(r)}$ and is concentrated within a spacetime region of size
$r$.

The effect of a single spacetime fluctuation can be described in
the path integral approach by adding a contribution $\int
\!{\cal D}\phi e^{-I_0} I_{\rm int}$ to the bare low-energy
Euclidean path integral $\int {\cal D}\phi e^{-I_0}$, $I_0$
being the bare low-energy action.  If we consider ${\cal N}$
indistinguishable gravitational fluctuations, the contribution
is $\int \!{\cal D}\phi e^{-I_0} (I_{\rm int})^{\cal N}/{\cal
N}!$. Thus, summing over any number ${\cal N}$ of them, we
obtain the path integral $\int\! {\cal D}\phi e^{-I_0+I_{\rm
int}}$.

The bilocal effective action above does not lead to a unitary
evolution for the low-energy fields because there exist different
trajectories that arrive at a given configuration
$(\phi,\dot\phi)$; the future evolution depends on these past
trajectories and not only on the values of $\phi$ and $\dot \phi$
at that instant of time. Therefore, it is not sufficient to know
the fields and their time derivatives at an instant of time in
order to know their values at a later time: we need to know the
history of the system, at least for a time $r$. As a consequence,
the system cannot possess a well-defined Hamiltonian vector field
and undergoes an intrinsic loss of predictability \cite{89eli01}.

The exponential of the interaction term $e^{I_{\rm int}}$ can be
written as \cite{96zin01}
\begin{equation}
\int \!{\cal D}\alpha
e^{-\frac{1}{2} \int \!dt dt^\prime
\gamma_{ij}(t-t^\prime)\alpha^i(t)\alpha^j(t^\prime)}
 e^{-\int\! dt \alpha^i(t) h_i(t)},
\end{equation}
where, the continuous matrix $\gamma_{ij}(t-t^\prime)$ is the
inverse of $c^{ij}(t-t^\prime)$, i.e.,
\begin{equation}
\int \!dt^{\prime\prime}\gamma_{ik}(t-t^{\prime\prime})
c^{kj}(t^{\prime\prime}-t^\prime)=\delta_i^j\delta(t-t^\prime).
\end{equation}
Note that the quadratic character of the distribution for the
fields $\alpha^i$ is a consequence of the weak-coupling
approximation (second order in $\epsilon$), which keeps only the
bilocal term in the action. Beyond the weak-coupling
approximation, higher-order terms would introduce deviations from
this noise distribution. Note also that we have a different field
$\alpha^i$ for each kind of interaction $h_i$. Thus, we have
transferred the nonlocality of the low-energy fields $\phi$ to
the set of fields $\alpha^i$, which are nontrivially coupled to
it.

If we now perform a Wick rotation back to Lorentzian spacetime,
we see that the path integral  has the form
\begin{equation}
\int\! {\cal D}\alpha P[\alpha]
\int \!{\cal D}\phi e^{i\big[S_0+
\int\! dt \alpha^i(t)h_i(t)\big]},
\end{equation}
where $S_0$ is the low-energy Lorentzian action and
\begin{equation}
P[\alpha]=e^{-\frac{1}{2} \int \!dt dt^\prime
\gamma_{ij}(t-t^\prime)\alpha^i(t)\alpha^j(t^\prime)}
 e^{-\int\! dt \alpha^i(t) h_i(t)}
\end{equation}
is the Gaussian probability distribution with correlation
functions $c^{ij}(t-t')$ for the stochastic nonlocal fields
$\alpha^i$ that represent spacetime foam and which are not
affected by the Wick rotation.

\section{Classical diffusion}
\label{class}

The analysis of the previous section ignores in a way the quantum
nature of gravitational fluctuations such as virtual black holes
or quantum time machines. Indeed, the fields $\alpha^i$ represent
quantum gravitational spacetime foam but, as we have seen, the
path integral for the whole system does not contain any trace of
the dynamical character of the fields $\alpha^i$. It just
contains a Gaussian probability distribution for them. The path
integral above can then be interpreted as a Gaussian average over
the classical noise sources $\alpha^i$. Classicality here means
that we can keep the sources $\alpha^i$ fixed, ignoring the noise
commutation relations in a kind of zeroth-order semiclassical
approximation, and, at the end of the calculations, we just
average over them. The next section will be devoted to the
quantum noise effects generated by spacetime foam that we are
ignoring here.

The master equation that governs the Lorentzian dynamics of the
low-energy fields in foamlike spacetimes is derived in what
follows. For each fixed set of fields $\alpha^i$, the evolution
equation for the density matrix $\rho_\alpha(t)$, obtained with
the Hamiltonian
\begin{equation}
H_\alpha(t)= H_0+ \alpha^i(t) h_i,
\end{equation}
$H_0$ being the bare Hamiltonian of the low-energy field,
is
\begin{equation}
\dot \rho_\alpha(t)=-i[H_0,\rho_\alpha(t)]
-i\alpha^i(t)[h_i,\rho_\alpha(t)].
\end{equation}
In the interaction picture, this equation becomes
\begin{equation}
\dot \rho^{\sc i}_\alpha(t)=
-i\alpha^i(t)[h_i^{\sc i}(t),\rho^{\sc i}_\alpha(t)],
\end{equation}
where
\begin{eqnarray}
\rho^{\sc i}_\alpha(t)&=&U^+_0(t)\rho_\alpha(t)U_0(t),
\\
h_i^{\sc i}(t)&=&U_0^+(t)h_iU_0(t)
\end{eqnarray}
with $U_0(t)=e^{-iH_0t}$. Averaging this equation over
$\alpha^i$ would provide a master equation for the density
matrix of the system although it would not be very useful since
it would contain terms of all orders in $\alpha^i$. To avoid
this problem, we integrate this equation between an initial time
$t_0$ and $t$:
\begin{equation}
\rho^{\sc i}_\alpha(t)=
\rho^{\sc i}_\alpha(t_0)-i\int_{t_0}^t \!\! dt'\alpha^i(t')
[h_i^{\sc i}(t'),\rho_\alpha^{\sc i}(t')]
\end{equation}
and introduce this formal solution back to the differential
equation for $\rho_\alpha^{\sc i}$ noting that at the initial time
$\rho_\alpha^{\sc i}(t_0)=\rho^{\sc i}(t_0)$ does not depend on
$\alpha^i$ (if this were not the case, there would be an extra
renormalization term in the Hamiltonian):
\begin{eqnarray}
\dot \rho_\alpha^{\sc i}(t) &=& -i\alpha^i(t)[h_i^{\sc i}(t),
\rho^{\sc i}(t_0)]
\nonumber\\
&&-\int_{t_0}^t \!\! dt' \alpha^i(t)\alpha^j(t')
[h_i^{\sc i}(t),[h_j^{\sc i}(t'),\rho_\alpha^{\sc i}(t')]].
\end{eqnarray}

Next, we perform the Gaussian average over $\alpha^i$ taking into
account that $\rho_\alpha^{\sc i}(t)$ does not depend on
$\alpha^i$ at zeroth order but only at first order, i.e.,
$\rho_\alpha^{\sc i}(t)=\rho^{\sc i}(t) +O(\alpha)$ with
$\rho^{\sc i}(t)=\langle \rho_\alpha^{\sc i}(t)\rangle$ and keep
terms up to second order in $\epsilon$ (weak-coupling
approximation). We then obtain the following equation for
$\rho^{\sc i}(t)$:
\begin{eqnarray}
\dot \rho^{\sc i}(t) =&&-\int_0^{t-t_0} \!\!\!\!\!\!
d\tau \langle\alpha^i(t)
\alpha^j(t-\tau) \rangle
\nonumber
\\
&&\times[h_i^{\sc i}(t),[h_j^{\sc i}(t-\tau),
\rho^{\sc i}(t-\tau)]],
\end{eqnarray}
where we have made a change of integration variables from $t'$ to
$\tau=t-t'$. We also assume that $\rho^{\sc i}(t)$ hardly changes
within a correlation time $r$ (Markov approximation), so that
$\rho^{\sc i}(t-r) \sim \rho^{\sc i}(t)$. This amounts to ignore
terms of order $\epsilon^4$ in the master equation. The initial
condition can be taken at $t_0=-\infty$, so that the integration
range is now $(0,\infty)$. Note that, since $ \langle\alpha^i(t)
\alpha^j(t-\tau) \rangle=c^{ij}(\tau)$ is nonvanishing only for
$\tau<r$, this limit $t_0\rightarrow -\infty$ just implies that
the evolution must take place over periods of time much larger
than the correlation time $r$ for the approximation to be valid.
The resulting master equation in the interaction picture is then
\begin{equation}
\dot \rho^{\sc i}(t) =-\int_0^\infty \!\!\! d\tau c^{ij}(\tau)
[h_i^{\sc i}(t),[h_j^{\sc i}(t-\tau), \rho^{\sc i}(t)]],
\end{equation}

Transforming this equation back to the Schr\"{o}dinger picture, we
obtain the equation
\begin{equation}
\dot \rho= -i [ H_0,\rho]-
\int_0^\infty\!\!\! d\tau c^{ij}(\tau)
\left[h_{i},\left[h_{j}^{\sc i}(-\tau),\rho\right]\right].
\end{equation}
Since $h_j^{\sc i}(-\tau)=U_0^+(-\tau)h_jU_0(-\tau)$ and
$U_0(\tau)=1+O(\tau/l)$, the final form of the master equation
for a low-energy system subject to gravitational fluctuations
treated as a classical environment and  at zeroth order in $r/l$
(the effect of higher order terms in $r/l$ will be thoroughly
studied in the next section) is \cite{84ban01}
\begin{equation}
\dot \rho= -i [ H_0,\rho]-
\int_0^\infty \!\!\! d\tau c^{ij}(\tau)
\left[h_{i},\left[h_{j},\rho\right]\right].
\end{equation}

The first term would also be present in the absence of
fluctuations, since it governs the low-energy Hamiltonian
evolution. The second term is a direct consequence of the
foamlike structure of spacetime and the related existence of a
minimum length. It is a diffusion term which will be responsible
for the loss of coherence. Note that a dissipation term,
necessary to preserve the commutation relations under time
evolution, is not present. However, we have considered the
classical noise limit, i.e., the fields $\alpha^i$ have been
considered as classical sources and the commutation relations are
automatically preserved. We will see that the dissipation term,
apart form being of quantum origin, is $r/l$ times smaller than
the diffusion term and we have only considered the zeroth order
approximation in $r/l$.

The diffusion term induces a characteristic decoherence time
$\tau_{d}$ that can be easily calculated. Indeed, the interaction
Hamiltonian density $h_i$ is of order
$l_*^{-4}(l_*/l)^{2n_{\underline{i}}(1+s_{\underline{i}})}$ and
$c^{ij}(\tau)$ is of order $e^{-S(r)}$. Furthermore, the
diffusion term contains one integral over time and two integrals
over spatial positions. The integral over time and the one over
relative spatial positions provide a factor $r^4$, since
$c^{ij}(\tau)$ is different from zero only in a spacetime region
of size $r^4$, and the remaining integral over global spatial
positions provides a factor $l^3$, the typical low-energy spatial
volume. Putting everything together, we see that the diffusion
term is of order
$l^{-1}\epsilon^2\sum_{{\underline{i}}{\underline{j}}}
(l_*/l)^{\eta_{\underline{i}}+\eta_{\underline{j}}}$, with
$\eta_{\underline{i}}=2n_{\underline{i}}(1+s_{\underline{i}})-2$.
This quantity defines the inverse of the decoherence time
$\tau_d$. Therefore, the ratio between the decoherence time
$\tau_d$ and the low-energy length scale $l$ is
\begin{equation}
\tau_d/l\sim \epsilon^{-2}\big[
\sum_{{\underline{i}}{\underline{j}}}
(l_*/l)^{\eta_{\underline{i}}+\eta_{\underline{j}}}\big]^{-1}.
\end{equation}
Only gravitational fluctuations whose size is very close to
Planck length will give a sufficiently small decoherence time,
because of the exponential dependence of $\epsilon\sim
e^{-S(r)/2} (r/l_*)^2$. Slightly larger fluctuations will have a
very small effect on the unitarity of the effective theory. For
the interaction term that corresponds to the mass of a scalar
field, the parameter $\eta$ vanishes and, consequently,
$\tau_d/l\sim \epsilon^{-2}$. Thus, the scalar mass term will
lose coherence faster than any other interaction. Indeed, for
higher spins and/or powers of the field strength, $\eta\geq 1$
and therefore $\tau_d/l$ increases by powers of $l/l_*$. For
instance, the scalar-fermion interaction term
$\phi^2\bar\psi\psi$, which has the next relevant decoherence
time, corresponds to a decoherence ratio $\tau_d/l\sim
\epsilon^{-2}l/l_*$. We see that the decoherence time for the
mass of scalars is independent of the low-energy length scale
and, for gravitational fluctuations of size close to Planck
length, $\epsilon$ may be not too small so that scalar masses may
lose coherence fairly fast, maybe in a few times the typical
evolution scale. Higher power and/or spin interactions will lose
coherence much slower but for sufficiently high energies
$l^{-1}$, although much smaller than the gravitational
fluctuations energy $r^{-1}$, the decoherence time may be small
enough. This means that quantum fields will lose coherence faster
for higher-energy regimes.

\section{Quantum bath}
\label{qme}

In this section, we will take into account the quantum dynamical
character of the fields $\alpha^i$ that represent spacetime
gravitational fluctuations (e.g., virtual black holes or quantum
time machines) and describe spacetime foam in terms of a quantum
thermal bath. By comparing the system consisting of low-energy
fields suitably coupled to a quantum bath \cite{91gar01,car93}
with the results obtained above for gravitational fluctuations,
we will see that spacetime foam can be substituted by an
effective quantum thermal bath.

Let us start studying a system with a Hamiltonian
\begin{equation}
H=H_0+H_{\rm int}+H_{\rm b},
\end{equation}
where $H_0$ is the bare Hamiltonian that represents the
low-energy fields and $H_{\rm b}$ is the Hamiltonian of a bath
that, for simplicity, will be represented by a real massless
scalar field. The interaction Hamiltonian will be chosen to have
the form $H_{\rm int}=\xi^i h_i$, the noise operators
$\xi^i$ being given by
\begin{equation}
\xi^{{\underline{i}}}(x,t)=i\int \! \frac{dk}{\sqrt \omega}
\chi^{{\underline{i}}}(\omega)
[ a^+(k) e^{i(\omega t-k x)}-
a(k) e^{-i(\omega t-k x)}].
\end{equation}
In this expression, $a$ and $a^+$ are, respectively, the
annihilation and creation operators associated with the bath,
$\omega=\sqrt{k^2}$, and $\chi^{{\underline{i}}}(\omega)$ are real
functions that represent the coupling between the system and the
bath for each frequency $\omega$ and for each interaction $h_i$.
These couplings $\chi^{{\underline{i}}}(\omega)$ can also be
written in the position representation if we note that the
momentum of the bath scalar field $p(x,t)$ has the form
\begin{equation}
p(x,t)=i\int\! dk \sqrt \omega [ a^+(k)
e^{i(\omega t-k x)}-a(k) e^{-i(\omega t-k x)}],
\end{equation}
so that the noise operators $\xi^i$ have the form
\begin{equation}
\xi^{\underline{i}}(x,t)= \int\! dx'
\chi^{{\underline{i}}}(x-x')p(x',t).
\end{equation}
Here,
\begin{equation}
\chi^{\underline{i}}(y)=\int\! \frac{dk}{\sqrt \omega}
\chi^{{\underline{i}}}(\omega) \cos (ky)
\end{equation}
represent the couplings between the low-energy field and the bath
in the position representation. Since we are trying to construct
a model for spacetime foam, we will assume that the couplings
$\chi^{{\underline{i}}}(y)$ will be concentrated on a region of
radius $r$ and therefore the couplings
$\chi^{{\underline{i}}}(\omega)$ will induce a significant
interaction with all the bath frequencies $\omega$ up to the
natural cutoff $r^{-1}$. Furthermore, these couplings have
dimensions of length and we will also assume that they are of
order $e^{-S(r)/2} r$. All the relevant information about the
couplings is encoded in the commutation relations and the
correlation function of the noise operators $\xi^i$.

Let us start with the commutation relations at different times
of the noise variables. Taking into account the commutation
relations for the annihilation and creation operators $a$ and
$a^+$, i.e.,
\begin{eqnarray}
&[a(k),a(k')]=[a^+(k),a^+(k')]=0,&
\\
&[a(k),a^+(k')]=\delta(k-k'),&
\end{eqnarray}
 it is easy to see that
\begin{equation}
[\xi^i(t),\xi^j(t')]=i \dot f^{ij}(t-t'),
\end{equation}
where
\begin{eqnarray}
f^{ij} (\tau)&=&\int_0^\infty \!\!\!
d\omega G^{ij}(\omega) \cos(\omega\tau),
\\
G^{ij}(\omega)&=&8\pi
\frac{\sin(\omega |x_{\underline{i}}-x_{\underline{j}}|)}
{\omega |x_{\underline{i}}-x_{\underline{j}}|}
\chi^{\underline{i}}(\omega)\chi^{\underline{j}}(\omega).
\end{eqnarray}
Note that the functions $G^{ij}(\omega)$ and, hence,
$f^{ij}(\tau)$ depend on the relative spatial distance
$|x_{\underline{i}}-x_{\underline{j}}|$, are symmetric in the
pair of indices $ij$ and are uniquely determined by the couplings
$\chi^{\underline{i}}(\omega)$ and vice versa. In particular,
they are completely independent of the state of the bath or the
system.

In order to compare this model with that of topological
fluctuations previously described, it is convenient to introduce
the so-called commutative noise representation \cite{91gar01} by
defining new noise operators $ \alpha^i$ in the following form:
\begin{equation}
\alpha^i(t)Q(t^\prime)\equiv
\frac{1}{2}[\xi^i(t),Q(t^\prime)]_+
\end{equation}
for any operator $Q$. As we have seen, the commutators of the
noise operators $\xi^i$ at different times are $c$-numbers.
Therefore, it is straightforward to check that the operators $
\alpha^i$ commute at any time, i.e.,
\begin{equation}
[ \alpha^i(t), \alpha^j(t')]=0.
\end{equation}
However, the commutator of $\alpha^i$ with any low-energy
operator $A$ is in general nonvanishing and has the form:
\begin{equation}
[A(t), \alpha^i(t')]=\int_{0}^t \!\! d \tau
[A(t),h_j(\tau)]
\dot f^{ij} (t^\prime-\tau),
\end{equation}
with $h_i(t)={\cal U}^+(t)h_i{\cal U}(t)$ and ${\cal
U}(t)=e^{-iHt}$. The function $f^{ij}(\tau)$ can be interpreted
as a kind of memory function. Indeed, these commutators are
nonzero for low-energy operators that are in the future or, at
most, in the near past of the noise and vanish only when they are
in the far past. Only in the so-called first Markov approximation
the frontier among both regimes is sharply located where both
noise and low-energy fields are at the same instant of time.

We are now ready, following similar steps to those outlined in
the previous section, to write down the master equation for the
low-energy density matrix.  We will describe the whole system
(low-energy field and bath) by a density matrix $\rho_{\sc t}(t)$.
We will assume that, initially, the low energy fields and the
bath are independent, i.e., that at the time $t_0$
\begin{equation}
\rho_{\sc t}(t_0)=\rho(t_0)\otimes \rho_{\rm b}.
\end{equation}
As in the classical noise case, if the low-energy fields and the
bath do not decouple at any time, an extra renormalization term
should be added to the Hamiltonian.  In the interaction picture,
the density matrix has the form
\begin{equation}
\rho_{\sc t}^{\sc i}(t)=U^+(t)\rho_{\sc t}(t)U(t),
\end{equation}
with $U(t)=U_0(t)U_{\rm b}(t)$, where $U_0(t)= e^{-iH_0t}$ and
$U_{\rm b}(t)=e^{-iH_{\rm b}t}$. It obeys the equation of motion
\begin{equation}
\dot\rho_{\sc t}^{\sc i}(t)=-i [\xi^i(t)
h_i^{\sc i}(t),\rho_{\sc t}^{\sc i}(t)].
\end{equation}
Here,
\begin{eqnarray}
\xi^i(t)&=&U^+(t)\xi^iU(t)=U_{\rm b}^+(t)\xi^iU_{\rm b}(t),
\\
h_i^{\sc i}(t)&=&U^+(t)h_iU(t)=U_0^+(t)h_iU_0(t).
\end{eqnarray}
Integrating this evolution equation and introducing the result
back into it, we obtain the following integro-differential
equation:
\begin{eqnarray}
\dot\rho_{\sc t}^{\sc i}(t)&=&-i[\xi^i(t)
h_i^{\sc i}(t),\rho_{\sc t}^{\sc i}(t_0)]
\nonumber\\
&&-\int_{t_0}^t \!\!dt'[\xi^i(t) h_i^{\sc i}(t),
[\xi^j(t') h_j^{\sc i}(t'),\rho_{\sc t}^{\sc i}(t')]].
\end{eqnarray}
If we now trace over the variables of the bath, define $\rho^{\sc
i}(t)\equiv{\rm tr}_{\rm b} [\rho_{\sc t}^{\sc i}(t)]$ and note
that ${\rm tr}_{\rm b}[\xi^i(t)h_i^{\sc i}(t)\rho_{\sc t}^{\sc
i}(t_0)]=0$ (because ${\rm tr}_{\rm b}[\xi^i(t)\rho_{\rm b}]=0$),
we obtain
\begin{equation}
\dot\rho^{\sc i}(t)=
-\int_{t_0}^t \!\! dt' {\rm tr}_{\rm b}
\left\{[\xi^i(t) h_i^{\sc i}(t),
[\xi^j(t') h_j^{\sc i}(t'),
\rho_{\sc t}^{\sc i}(t')]]\right\}.
\end{equation}

In the weak-coupling approximation, which implies that $\xi^ih_i$
is much smaller than $H_0$ and $H_{\rm b}$ (this is justified
since it is of order $\epsilon$), we assume that the bath density
matrix does not change because of the interaction, so that
$\rho_{\sc t}^{\sc i}(t)=\rho^{\sc i}(t)\otimes\rho_{\rm b}$. The
error introduced by this substitution is of order $\epsilon$ and
ignoring it in the master equation amounts to keep terms only up
to second order in this parameter. Since $[\xi^i(t), h_j^{\sc
i}(t')]=0$ because $[\xi^i, h_j]=0$, the right hand side of this
equation can be written in the following way
\begin{eqnarray}
-\frac{1}{2}\int_{t_0}^t \!\!dt'&&\left\{
\langle[\xi^i(t),\xi^j(t')]_+\rangle
[h_i^{\sc i}(t),[h_j^{\sc i}(t'),\rho^{\sc i}(t')]]\right.
\nonumber\\
&&\left.+\langle[\xi^i(t),\xi^j(t')]\rangle
[h_i^{\sc i}(t),[h_j^{\sc i}(t'),\rho^{\sc i}(t')]_+]\right\},
\end{eqnarray}
where the average of any operator $Q$ has been defined as
$\langle Q\rangle\equiv {\rm tr}_{\rm b}(Q\rho_{\rm b})$. Next we
note that $\langle[\xi^i(t),\xi^j(t')]\rangle=i\dot f^{ij}(t-t')$
and, using the commutative noise representation introduced above,
we can write
\begin{equation}
\frac{1}{2}\langle[\xi^i(t),\xi^j(t')]_+\rangle=
\langle  \alpha^i(t) \alpha^j(t')\rangle\equiv
c^{ij}(t-t').
\end{equation}

If we make the assumption that the bath is in a thermal state
$\rho_{\rm b}={\cal Z}^{-1}e^{-H_{\rm b}/T}$ with a temperature
inversely proportional to the size of the gravitational
fluctuations (e.g., the radius of the virtual black holes, or the
size of the regions containing closed timelike curves in  the
case of quantum time machines), $T\sim 1/r$, the correlation
function $c^{ij}(t-t')$ acquires the form:
\begin{equation}
  c^{ij}(\tau)
=\int_0^\infty \!\!\! d\omega \omega
G^{ij}(\omega)[ N(\omega)+ 1/2]
\cos(\omega\tau),
\end{equation}
where $N(\omega)=\left[\exp(\omega/T)-1\right]^{-1}$ is the mean
occupation number of the bath corresponding to the frequency
$\omega$. In this calculation, we have made use of the following
relations, valid for a thermal state:
\begin{eqnarray}
&\langle a(k)\rangle=\langle a^+(k)\rangle=0,&
\\
&\langle a(k)a(k')\rangle=\langle a^+(k)a^+(k')\rangle=0,&
\\
&\langle a^+(k)a(k')\rangle=N(\omega)\delta(k-k').&
\end{eqnarray}
Similarly, we can easily compute the higher order correlations
$\langle  \alpha^i(t)  \alpha^j(t') \alpha^k(t'')\rangle$, etc.
Those containing an odd number of fields $\alpha^i$ turn out to
be identically zero while those containing an even number can be
written in terms of the two-point correlation function
$c^{ij}(\tau)$. This means that the trace $\langle Q\rangle$
corresponds to a Gaussian average over $\alpha^i$, provided that
the bath is in a thermal state, as we are
considering. In this way, we have established a relation between
a quantum thermal bath and spacetime foam, which can also be
described by a Gaussian average, as we have seen.

The Markov approximation allows the substitution of $\rho^{\sc
i}(t')$ by $\rho^{\sc i}(t)$ in the master equation because the
integral over $t'$ will get a significant contribution from times
$t'$ that are close to $t$ due to the factors $\dot f^{ij}(t-t')$
and $c^{ij}(t-t')$ and because, in this interval of time, the
density matrix $\rho^{\sc i}$ will not change significantly.
Indeed, the typical evolution time of $\rho^{\sc i}$ is the
low-energy time scale $l$, which will be much larger than the
time scale $r$ associated with the bath. If we perform a change
of the integration variable from $t'$ to $\tau=t-t'$, write
\begin{equation}
\rho^{\sc i}(t')=\rho^{\sc i}(t-\tau)=
\rho^{\sc i}(t)-\tau\dot\rho^{\sc i}(t) +O(\tau^2),
\end{equation}
and introduce this expression in the master equation above, we
easily see that the error introduced by the Markovian
approximation is of order $\epsilon^2$, i.e., it amounts ignore
a term of order $\epsilon^4$.  The upper integration limit $t$
in both integrals can be substituted by $\infty$ for evolution
times $t-t_0$ much larger than the correlation time $r$, because
of  the factors $\dot f^{ij}(\tau)$ and $c^{ij}(\tau)$ that
vanish for $\tau>r$, which is equivalent to take the initial
condition to the infinite past $t_0\rightarrow -\infty$.

Then, the master equation in the interaction picture acquires
the form
\begin{eqnarray}
\dot\rho^{\sc i}(t)&=& -\frac{i}{2}\int_0^\infty \!\!\! d\tau
\dot f^{ij}(\tau)
[h_i^{\sc i}(t),[h_j^{\sc i}(t-\tau),\rho^{\sc i}(t)]_+]
\nonumber\\
&&-\int_0^\infty \!\!\! d\tau
  c^{ij}(\tau)
[h_i^{\sc i}(t),[h_j^{\sc i}(t-\tau),\rho^{\sc i}(t)]].
\end{eqnarray}
We can now transform the resulting equation back to the
Schr\"{o}dinger picture
\begin{eqnarray}
\dot\rho&=& -i[H_0,\rho]-\frac{i}{2}\int_0^\infty \!\!\! d\tau
\dot f^{ij}(\tau) [h_i,[h_j^{\sc i}(-\tau),\rho]_+]
\nonumber\\
&&-\int_0^\infty \!\!\! d\tau
  c^{ij}(\tau) [h_i,[h_j^{\sc i}(-\tau),\rho]].
\end{eqnarray}
After an integration by parts, the second term of the right hand
side becomes
\begin{equation}
\frac{i}{2}f^{ij}(0)[h_ih_j,\rho]
-\frac{i}{2}\int_0^\infty \!\!\! d\tau  f^{ij}(\tau)
[h_i,[\dot h_j^{\sc i}(-\tau),\rho]_+].
\end{equation}
The first term is just a finite renormalization of the original
low-energy Hamiltonian from $H_0$ to
\begin{equation}
H_0'=H_0-\frac{1}{2}f^{ij}(0) h_ih_j
\end{equation}
and the master equation can then be written in its final form
\begin{eqnarray}
\dot\rho&=& -i[H_0',\rho]-\frac{i}{2}\int_0^\infty \!\!\! d\tau
f^{ij}(\tau)
[h_i,[\dot h_j^{\sc i}(-\tau),\rho]_+]
\nonumber\\
&&-\int_0^\infty \!\!\! d\tau
  c^{ij}(\tau) [h_i,[h_j^{\sc i}(-\tau),\rho]].
\end{eqnarray}

Before discussing this equation in full detail, let us first
study the classical noise limit. With this aim, let us introduce
the parameter
\begin{equation}
\sigma=\int\! dk' [a(k),a^+(k')],
\end{equation}
which is equal to 1 for quantum noise and 0 for classical noise.
Then, the $f$-term is proportional to $\sigma$ and therefore
vanishes in the classical noise limit. The $c$-term also contains
a factor $\sigma$ but, in addition, $N(\omega)$ becomes $N(\sigma
\omega)$ when introducing the parameter $\sigma$. In the limit
$\sigma\rightarrow 0$, the term proportional to $1/2$ in
$c^{ij}(\tau)$ vanishes and the term proportional to $N(\sigma
\omega)$ acquires the value $c_{\rm
class}^{ij}(\tau)=Tf^{ij}(\tau)$. Also, the renormalization term
of the low-energy Hamiltonian vanishes in this limit. In this
way, we have arrived at the same master equation that we obtained
in the previous section. This is not surprising because the
origin of the $f$-term is precisely the noncommutativity of the
noise operators, i.e., its quantum nature, while the $c_{\rm
class}$-term actually contains the temperature effects. At zeroth
order in $r/l$, the master equation for classical noise then
acquires the form
\begin{equation}
\dot \rho= -i [ H_0,\rho]-
\int_0^\infty \!\!\! d\tau c_{\rm class}^{ij}(\tau)
\left[h_{i},\left[h_{j},\rho\right]\right].
\end{equation}

Let us now analyze the general master equation, valid up to
second order in $\epsilon$ that takes into account the quantum
nature of the gravitational fluctuations. These contributions,
although small in the low-energy regime, might still be
experimentally testable. In addition, they may provide
interesting information about the higher-energy regimes in which
$l$ may be of the order of a few Planck lengths and for which the
weak-coupling approximation is still valid. In order to see these
contributions explicitly, let us further elaborate the master
equation. In terms of the operator $L_0$ defined as
$L_0A=[H_0,A]$ acting of any low-energy operator $A$, the time
dependent interaction $h^{\sc i}_j(-\tau)$ can be written as
\begin{equation}
h^{\sc i}_j(-\tau)=e^{-iL_0\tau}h_j.
\end{equation}
The interaction $h_j$ can be expanded in eigenoperators
$h_{j\Omega}^{\pm}$ of the operator $L_0$, i.e.,
\begin{equation}
h_j=\int \! d\mu_\Omega\left(h_{j\Omega}^++h_{j\Omega}^-\right),
\end{equation}
with $L_0h_{j\Omega}^{\pm}=\pm \Omega h_{j\Omega}^{\pm}$ and
$d\mu_\Omega$ being an appropriate  spectral measure, which is
naturally cut off around the low-energy scale $l^{-1}$.  This
expansion always exists provided that the eigenstates of $H_0$
form a complete set. Then, $h^{\sc i}_j(-\tau)$ can be written as
\begin{equation}
h^{\sc i}_j(-\tau)=\int \! d\mu_\Omega (e^{-i\Omega \tau }
h^+_{j\Omega }+e^{i\Omega \tau } h^-_{j\Omega }).
\end{equation}
It is also convenient to define the new interaction operators
for each low-energy frequency $\Omega$
\begin{eqnarray}
h^1_{j\Omega}&=&h^+_{j\Omega }-h^-_{j\Omega },
\\
h^2_{j\Omega}&=&h^+_{j\Omega }+h^-_{j\Omega }.
\end{eqnarray}

Both the term proportional to $f^{ij}(\tau)$ and the term
proportional to $c^{ij}(\tau)$ are integrated over $\tau\in
(0,\infty)$. Because of these incomplete integrals, each term
provides two different kinds of contributions coming from the
bulk term and the principal part in the well-know formula
\begin{equation}
\int_0^\infty \!\!\!d\tau e^{i\omega\tau}
=\pi\delta(\omega)+{\cal P}(i/\omega),
\end{equation}
where ${\cal P}$ is the Cauchy principal part \cite{cauchy}.

The master equation can then be written in the following form
\begin{equation}
\dot \rho =-(iL_0'+L_{\rm diss}+L_{\rm diff}+
iL_{\rm stark}+iL_{\rm lamb})\rho,
\end{equation}
where the meaning of the different terms are explained in what
follows.

The first term $-iL_0'\rho$, with $L_0'\rho=[H_0',\rho]$, is
responsible for the renormalized low-energy Hamiltonian
evolution. The renormalization term is of order $\varepsilon^2$
as compared with the low-energy Hamiltonian $H_0$, where
$\varepsilon^2= \epsilon^2\sum_{{\underline{i}}{\underline{j}}}
(l_*/l)^{\eta_{\underline{i}}+\eta_{\underline{j}}}$ and,
remember, $\eta_{\underline{i}}= 2n_{\underline{i}}
(1+s_{\underline{i}})-2$ is a parameter specific to each kind of
interaction term $h_i$.

The dissipation term
\begin{equation}
L_{\rm diss}\rho=-\frac{\pi}{4}\int\! d\mu_\Omega \Omega
G^{ij}(\Omega) [h_i,[h^1_{j\Omega},\rho]_+]
\end{equation}
is necessary for the preservation in time of the low-energy
commutators in the presence of quantum noise. As we have seen, it
is proportional to the commutator between the noise creation and
annihilation operators associated with the effective bath that
represents spacetime foam and, therefore, vanishes in the
classical noise limit. Its size is of order $\varepsilon^2r/l^2$.

The diffusion process is governed by
\begin{equation}
L_{\rm diff}\rho=\frac{\pi}{2}\int \! d\mu_\Omega \Omega
G^{ij}(\Omega)[N(\Omega )+1/2] [h_i,[h^2_{j\Omega},\rho]],
\end{equation}
which contains two contributions: the first one is a temperature
effect of order $\varepsilon^2/l$ and the second is a cold
diffusion originated in the vacuum fluctuations of the
gravitational field and it is of order $\varepsilon^2r/l^2$. In
the classical noise limit, only the first contribution survives
and was already studied in the previous section.

The next term provides an energy shift which can be interpreted
as a gravitational ac Stark effect by comparison with its
quantum optical analog \cite{91gar01,car93}. Its expression is
\begin{equation}
L_{\rm stark}\rho =
\int\! d\mu_\Omega
{\cal P}\!\!\int_{0 }^{\infty }\!\!\! d\omega
\frac{\omega \Omega }{\omega^2- \Omega^2 }
G^{ij}(\omega)N(\omega )
[h_i,[h^2_{j\Omega},\rho]].
\end{equation}
Although it is also a temperature-dependent effect with the same
origin as the diffusion term, it contains a Cauchy principal
part. This translates into the fact that it is smaller than the
diffusion term although it does not vanish in the classical
noise limit. It is of order $\varepsilon^2r/l^2$.

Finally, $L_{\rm lamb}\rho$ is an energy shift generated by the
vacuum fluctuations of the gravitational  field (as the
dissipation term and the cold diffusion term) and that can
therefore  be interpreted as a gravitational Lamb shift. It has
the form
\begin{eqnarray}
L_{\rm lamb}\rho =&&\frac{1}{2}\int\! d\mu_\Omega
{\cal P}\!\!\int_{0 }^{\infty }\!\!\!
d\omega\frac{\Omega }{\omega^2- \Omega^2 }
G^{ij}(\omega)
\nonumber\\
&&\times \left\{\omega [h_i,[h^1_{j\Omega},\rho]]
-\Omega [h_i,[h^2_{j\Omega},\rho]_+]\right\}.
\end{eqnarray}
The second term is of order $\varepsilon^2r^2/l^3$, which is
fairly small. However, the first term will provide a significant
contribution of order $\varepsilon^2r/l^2\log(l/r)$. This
logarithmic dependence on the relative scale is indeed
characteristic of the Lamb shift \cite{91gar01,car93,itzy}.

As a summary, the $c$-term gives rise to four different
contributions: a thermal diffusion term, another diffusion term
originated from the vacuum fluctuations of the bath, a
contribution to what can be interpreted as a gravitational Lamb
shift, and, finally, a shift in the scalar-field oscillation
frequencies that can be interpreted as a gravitational Stark
effect. The $f$-term provides a dissipation part, necessary for
the preservation of commutators, and another contribution to the
gravitational Lamb shift. The size of these effects generated by
spacetime foam, compared with the bare evolution, are the
following: the thermal diffusion term is of order
$\varepsilon^2$, which is the only one that survived in the
approximations of the previous section; the diffusion created by
vacuum fluctuations, the damping term, and the Stark effect are
smaller by a factor $r/l$; and the Lamb shift has two
contributions: one is smaller than the diffusion term by a factor
$(r/l)^2$ and the other is of order $(r/l)\log(l/r)$ as compared
with the diffusion term. Note that the quantum effects induced by
spacetime foam become relevant as the low-energy length scale $l$
decreases, as we see from the fact that these effects depend on
the ratio $r/l$, while, in this situation, the diffusion process
becomes slower, except for the mass of scalars, which always
decoheres in a time scale which is close to the low-energy
evolution time.

These quantum gravitational effects could be measured, at least
in principle, since they are just energy shifts and decoherence
effects similar to those appearing in other areas of physics,
where fairly well established experimental procedures and results
exist, and which can indeed be applied here --- such as those
briefly discussed below --- provided that sufficiently high
accuracy can be achieved. On the other hand, scalar fields lose
quantum coherence extremely fast and Hawking has argued
\cite{96haw01} that this might be the reason for not observing
the Higgs particle. He has also suggested that loss of quantum
coherence might be responsible for the vanishing of the $\theta$
angle in quantum chromodynamics \cite{96haw01}.

Neutral kaon beams have been proposed as experimental systems for
measuring loss of coherence owing to quantum gravitational
fluctuations \cite{el84,hp95,bf98}. In these systems, the main
experimental consequence of the diffusion term (together with the
dissipative one necessary for reaching a stationary regime) is
violation of CPT \cite{pa82,unpred} because of the nonlocal
origin of the effective interactions. The estimates for this
violation are very close to the values accessible by current
experiments with neutral kaons and will be within the range of
near-future experiments. Macroscopic neutron interferometry
\cite{el84,ze84} provides another kind of experimental systems in
which the effects of the diffusion term have measurable
consequences since they may cause the disappearance of the
interference fringes \cite{el84,ze84}.

As for the gravitational Lamb and Stark effects, they are energy
shifts that depend on the frequency, so that different low-energy
modes will undergo different shifts. This translates into a
modification of the dispersion relations, which makes the
velocity of propagation frequency-dependent, as if low-energy
fields propagated in a ``medium''. Therefore, upon arrival at the
detector, low-energy modes will experience different time delays
(depending on their frequency) as compared to what could be
expected in the absence of quantum gravitational fluctuations.
These time delays in the detected signals will be very small
in general. However, it is still possible to measure them
 if we make the low-energy particles travel large
(cosmological) distances. In fact, $\gamma$-ray bursts provide
such a situation as has been recently pointed out \cite{am98},
thus opening a new doorway to observations of these quantum
gravitational effects. Indeed, the ratio between the time delay
owing to gravitational fluctuations and the width of the
intrinsic time structure of $\gamma$-ray bursts has been
estimated to be of order 1 for emissions with millisecond time
structure and energy around 20 MeV, provided that they travel a
distance of $10^{10}$ light years \cite{am98}, which are
compatible with $\gamma$-ray burst observations. If this
sensitivity can actually be reached, one would expect that the
presence of the gravitational Lamb and Stark shifts predicted
above could be observationally tested.

\section{Virtual black holes, wormholes, and time machines}
\label{compo}

It is well-known that it is not possible to classify all
four-dimensional topologies \cite{ha78} and, consequently, all
the possible components of spacetime foam. Here, we will briefly
discuss three different kinds of fluctuations: simply connected
nontrivial topologies, multiply connected topologies with trivial
second homology group (i.e. with vanishing second Betti number),
and finally spacetimes with a nontrivial causal structure, i.e.,
with closed timelike curves, in a bounded region.

The effective description proposed in this paper and the
associated master equation are particularly suited to the study
of low-energy effects produced by simply connected topology
fluctuations (e.g., virtual black holes). Hawking \cite{96haw01}
has shown that compact simply connected bubbles with the topology
$S^2\times S^2$ (whose second Betti number is $B_2=1$) can be
interpreted as closed loops of virtual black holes if one
realizes \cite{gibbons} that the process of creation of a pair of
real charged black holes accelerating away from each other in a
spacetime which is asymptotic to $\Re^4$ is provided by the Ernst
solution \cite{ernst}. This solution has the topology $S^2\times
S^2$ minus a point (which is sent to infinity) and this topology
is the topological sum of the bubble $S^2\times S^2$ plus
$\Re^4$. Virtual black holes will not obey classical equations of
motion but will appear as quantum fluctuations of spacetime and
thus will become part of the spacetime foam. Particles could fall
into these black holes and be re-emitted. The scattering
amplitudes of these processes \cite{96haw01} could be interpreted
as being produced by nonlocal effective interactions that would
take place inside the fluctuations and the master equation
obtained above could then be interpreted as providing the
evolution of the low-energy density matrix in the presence of a
bath of ubiquitous quantum topological fluctuations of the
virtual-black-hole type.

Wormholes \cite{haw88}, i.e., multiply connected fluctuations
(with vanishing second Betti number), also admit a description in
terms of nonlocal interactions that, in the weak-coupling
approximation, become bilocal. These quantum fluctuations connect
spacetime points that may be far apart from each other, in the
dilute gas approximation. Therefore, diffeomorphism invariance on
each spacetime region requires the coefficients $c^{ij}$ of this
bilocal interaction term to be spacetime independent. The same
conclusion can also be reached if we analyze wormholes from the
point of view of the universal covering manifold, which is, by
definition, simply connected. A wormhole is then represented in
the universal covering manifold by two boundaries, suitably
identified, located at infinity. This identification can be
implemented by introducing coefficients $c^{ij}$ that relate the
bases of the Hilbert space of wormholes in both regions of the
universal covering manifold. As coefficients in a change of
basis, $c^{ij}$ cannot depend on spacetime positions and,
therefore, will just be constant. This means that the fields
$\alpha^i$ cannot be interpreted as noise sources that are
Gaussian distributed at each spacetime point independently,
because the correlation time for the fields $\alpha^i$ is
infinite. Indeed, the constancy of $c^{ij}$ implies that they are
infinitely coherent and the Gaussian distribution to which they
are subject is therefore global, spacetime independent
\cite{col88}. One could still expect some effects originated in
the quantum nature of $\alpha^i$ such as a cold diffusion term in
the master equation or even dissipation. However, because they
are spacetime independent, they commute with every operator,
including low-energy ones, thus giving rise to superselection
sectors. Therefore, all the terms in the master equation, except
the one responsible for the unitary low-energy evolution, vanish.
Still, wormholes can be represented by a thermal bath as we have
done with localized gravitational fluctuations. However, in order
to reproduce their infinite correlation time, the couplings
$\xi^i$ between the bath and the low-energy fields must be
constant, they must commute with every other operator, and,
related to these two facts, only the zero-frequency (i.e.,
infinite wavelength) mode of the bath can be coupled to the
low-energy fields, thus leading to a unitary effective theory.

From the semiclassical point of view, most of the hitherto
proposed time machines \cite{morris} are unstable because quantum
vacuum fluctuations generate divergences in the stress-energy
tensor, i.e., are subject to the chronology protection conjecture
\cite{cpc}. However, quantum time machines \cite{pedro} confined
to small spacetime regions, for which the chronology protection
conjecture does not apply \cite{nocpc}, are likely to occur
within the realm of spacetime foam, where strong causality
violations or even the absence of a causal structure are
expected. These Planck-size regions with quantum time machines
admit an effective representation in terms of nonlocal
interactions that account for the causality violations and will
lead to a loss of quantum coherence \cite{ha95} that can also be
effectively described as coming from the interaction of
low-energy fields with a thermal bath. In this case, the
low-energy density matrix will also evolve according to the
master equation obtained in the previous sections.

\section{Conclusions}
\label{concl}

In this paper, we have built an effective theory in which quantum
gravitational spacetime foam has been substituted by a fixed
classical background plus nonlocal interactions between the
low-energy fields confined to bounded spacetime regions of nearly
Planck size. In the weak-coupling approximation, these nonlocal
interactions become bilocal. The low-energy evolution is not
unitary because of the absence of a nonvanishing timelike
Hamiltonian vector field. The nonunitarity of the bilocal
interaction can be encoded in a classical noise source locally
coupled to the low-energy fields and subject to a Gaussian
probability distribution. Then, the evolution of low-energy
fields is provided by a master equation which contains a
diffusion term. This diffusion is a direct consequence of the
nonlocal character of the quantum gravitational fluctuations
encompassed by spacetime foam. The decoherence rate is suppressed
by powers of the ratio between the gravitational fluctuation size
and the low-energy length scale, except for the mass interaction
term of scalar fields for which this rate is comparable with the
low-energy evolution scale.

We have argued that the quantum nature of spacetime foam is not
represented in this effective theory but only its thermal
properties. A model in terms of a quantum thermal field, which in
the classical noise limit coincides with the one described above,
has been proposed as describing the quantum and thermal
properties of spacetime foam. In this model, the low-energy
density matrix evolves according to a master equation that, apart
from inducing loss of coherence, contains additional terms that
may be relevant for sufficiently high energies. These terms
correspond to a dissipation process that ensure the preservation
of commutators, a cold diffusion, and energy shifts that can be
interpreted as gravitational Lamb and Stark effects. We have also
briefly discussed some of the possible experimental implications
that these quantum gravitational effects may have. A constructive
model in terms of nonlocal interactions that takes into account
the quantum origin of spacetime foam will be developed elsewhere
\cite{garayinprep,potsdam} within the formalism of Feynman and
Vernon \cite{feyver,feyver2,calde}.

Finally, among the possible components of spacetime foam, the
role of virtual black holes and small bounded regions that
contain closed timelike curves has been briefly analyzed in the
context of our effective model. We have also argued that dilute
wormholes do not admit a description in terms of a thermal bath
coupled to the whole low-energy spectrum and that they are
infinitely coherent as was already shown by Coleman \cite{col88}.

\acknowledgments

I am very grateful to G.A. Mena Marug\'{a}n, P.F.
Gonz\'{a}lez-D\'{\i}az, C. Barcel\'{o}, J.M. Raya, I.L. Egusquiza, C.
Cabrillo and J.I. Cirac for helpful discussions. I was supported
by funds provided by DGICYT and MEC (Spain) under Contract
Adjunct to the Project No. PB94--0107.

\end{document}